\documentclass[11pt]{article}
\usepackage{moriond,epsfig}

\def\be{\begin{equation}}
\def\ee{\end{equation}}
\def\bea{\begin{eqnarray}}
\def\eea{\end{eqnarray}}

\newcommand\aj{\ref@jnl{AJ}}%
\newcommand\araa{\ref@jnl{ARA\&A}}%
\newcommand\apj{{ApJ}}%
\newcommand\apjl{{ApJ}}%
\newcommand\apjs{\ref@jnl{ApJS}}%
\newcommand\ao{\ref@jnl{Appl.~Opt.}}%
\newcommand\apss{\ref@jnl{Ap\&SS}}%
\newcommand\aap{\ref@jnl{A\&A}}%
\newcommand\aapr{\ref@jnl{A\&A~Rev.}}%
\newcommand\aaps{\ref@jnl{A\&AS}}%
\newcommand\azh{\ref@jnl{AZh}}%
\newcommand\baas{\ref@jnl{BAAS}}%
\newcommand\jrasc{\ref@jnl{JRASC}}%
\newcommand\memras{\ref@jnl{MmRAS}}%
\newcommand\mnras{{MNRAS}}%
\newcommand\pra{\ref@jnl{Phys.~Rev.~A}}%
\newcommand\prb{\ref@jnl{Phys.~Rev.~B}}%
\newcommand\prc{\ref@jnl{Phys.~Rev.~C}}%
\newcommand\prd{\ref@jnl{Phys.~Rev.~D}}%
\newcommand\pre{\ref@jnl{Phys.~Rev.~E}}%
\newcommand\prl{\ref@jnl{Phys.~Rev.~Lett.}}%
\newcommand\pasp{\ref@jnl{PASP}}%
\newcommand\pasj{\ref@jnl{PASJ}}%
\newcommand\qjras{\ref@jnl{QJRAS}}%
\newcommand\skytel{\ref@jnl{S\&T}}%
\newcommand\solphys{\ref@jnl{Sol.~Phys.}}%
\newcommand\sovast{\ref@jnl{Soviet~Ast.}}%
\newcommand\ssr{\ref@jnl{Space~Sci.~Rev.}}%
\newcommand\zap{\ref@jnl{ZAp}}%
\newcommand\nat{\ref@jnl{Nature}}%
\newcommand\iaucirc{\ref@jnl{IAU~Circ.}}%
\newcommand\aplett{\ref@jnl{Astrophys.~Lett.}}%
\newcommand\apspr{\ref@jnl{Astrophys.~Space~Phys.~Res.}}%
\newcommand\bain{\ref@jnl{Bull.~Astron.~Inst.~Netherlands}}%
\newcommand\fcp{\ref@jnl{Fund.~Cosmic~Phys.}}%
\newcommand\gca{\ref@jnl{Geochim.~Cosmochim.~Acta}}%
\newcommand\grl{\ref@jnl{Geophys.~Res.~Lett.}}%
\newcommand\jcp{\ref@jnl{J.~Chem.~Phys.}}%
\newcommand\jgr{\ref@jnl{J.~Geophys.~Res.}}%
\newcommand\jqsrt{\ref@jnl{J.~Quant.~Spec.~Radiat.~Transf.}}%
\newcommand\memsai{\ref@jnl{Mem.~Soc.~Astron.~Italiana}}%
\newcommand\nphysa{\ref@jnl{Nucl.~Phys.~A}}%
\newcommand\physrep{\ref@jnl{Phys.~Rep.}}%
\newcommand\physscr{\ref@jnl{Phys.~Scr}}%
\newcommand\planss{\ref@jnl{Planet.~Space~Sci.}}%
\newcommand\procspie{\ref@jnl{Proc.~SPIE}}%

\newcommand{\chandra}{\emph{Chandra}}
\newcommand{\xmm}{XMM-\emph{Newton}}
\newcommand{\tspec}{$T_{\rm spec}$}
\newcommand{\tew}{$T_{\rm ew}$}
\newcommand{\tsl}{$T_{\rm sl}$}

\def\citealt{\cite}
\def\citep{\cite}
\def\citet{\cite}
\def\2A     {{2A 0335+096}}

\begin{document}
\vspace*{4cm}
\title{Comparing the temperatures of galaxy clusters from hydro-N-body
simulations to {\it Chandra} and XMM-{\it Newton} observations
}

\author{ P. Mazzotta$^{1}$,
 E. Rasia$^2$, L. Moscardini$^2$, G. Tormen$^2$  }

\address{$^1$Dip. di Fisica, Universit\`a di "Tor Vergata", via della Ricerca Scientifica 1, I-00133 Roma, Italy\\
$^2$Dipartimento di Astronomia, Universit\`a di Padova, vicolo
dell'Osservatorio 2, I-35122 Padova, Italy  
 }

\maketitle\abstracts{Theoretical studies of the physical processes in
 clusters of galaxies are mainly
based on the results of numerical  simulations,
which in turn are often directly compared to X-ray observations.
Although trivial in principle, these comparisons are not always
simple. We show that the projected spectroscopic temperature of
clusters obtained from X-ray observations is always
lower than the emission-weighed temperature.  This 
bias is related to the fact that the emission-weighted
temperature does not reflect the actual spectral properties of the
observed source. This has implications for the study of
thermal structures in clusters, especially when strong temperature
gradients, like shock fronts, are present.  In
real observations shock fronts appear much weaker than what is
predicted by emission-weighted temperature maps.
  We propose a new formula, the
spectroscopic-like temperature function that better approximates the spectroscopic
temperature, making simulations more directly
comparable to observations.}

\section{Introduction}
Recent  observational data with high spatial and spectral resolution suggest that clusters
are far from isothermal, and show instead a number of peculiar thermal
features, like cold fronts [\citealt{2000ApJ...541..542M},
\citealt{2001ApJ...555..205M}] cavities [\citealt{2000ApJ...534L.135M}], blobs and filaments [\citealt{2001MNRAS.321L..33F},\citealt{2003ApJ...596..190M}].

\begin{figure*}
\psfig{file=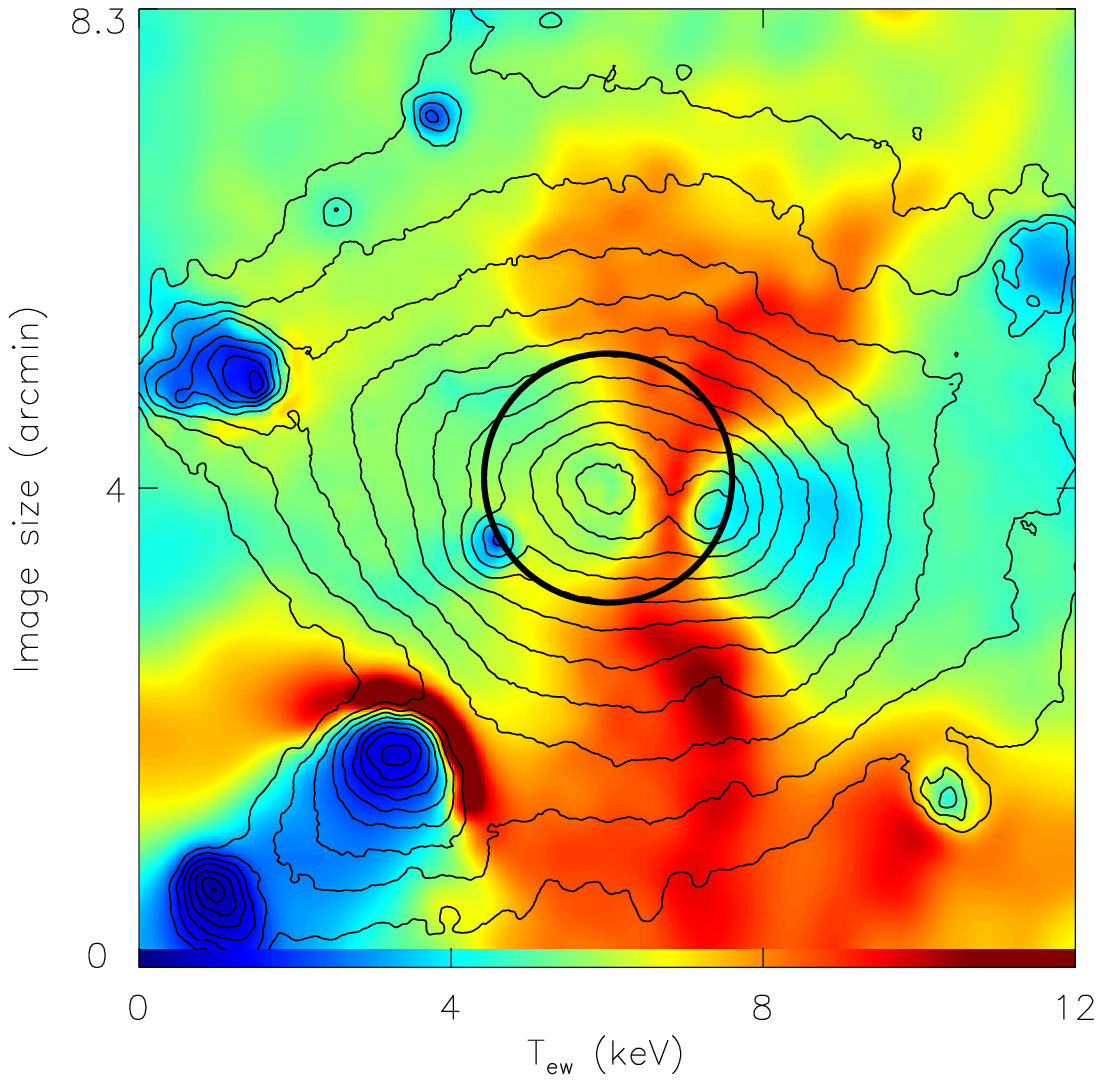,width=.25\textwidth}\hfil \psfig{file=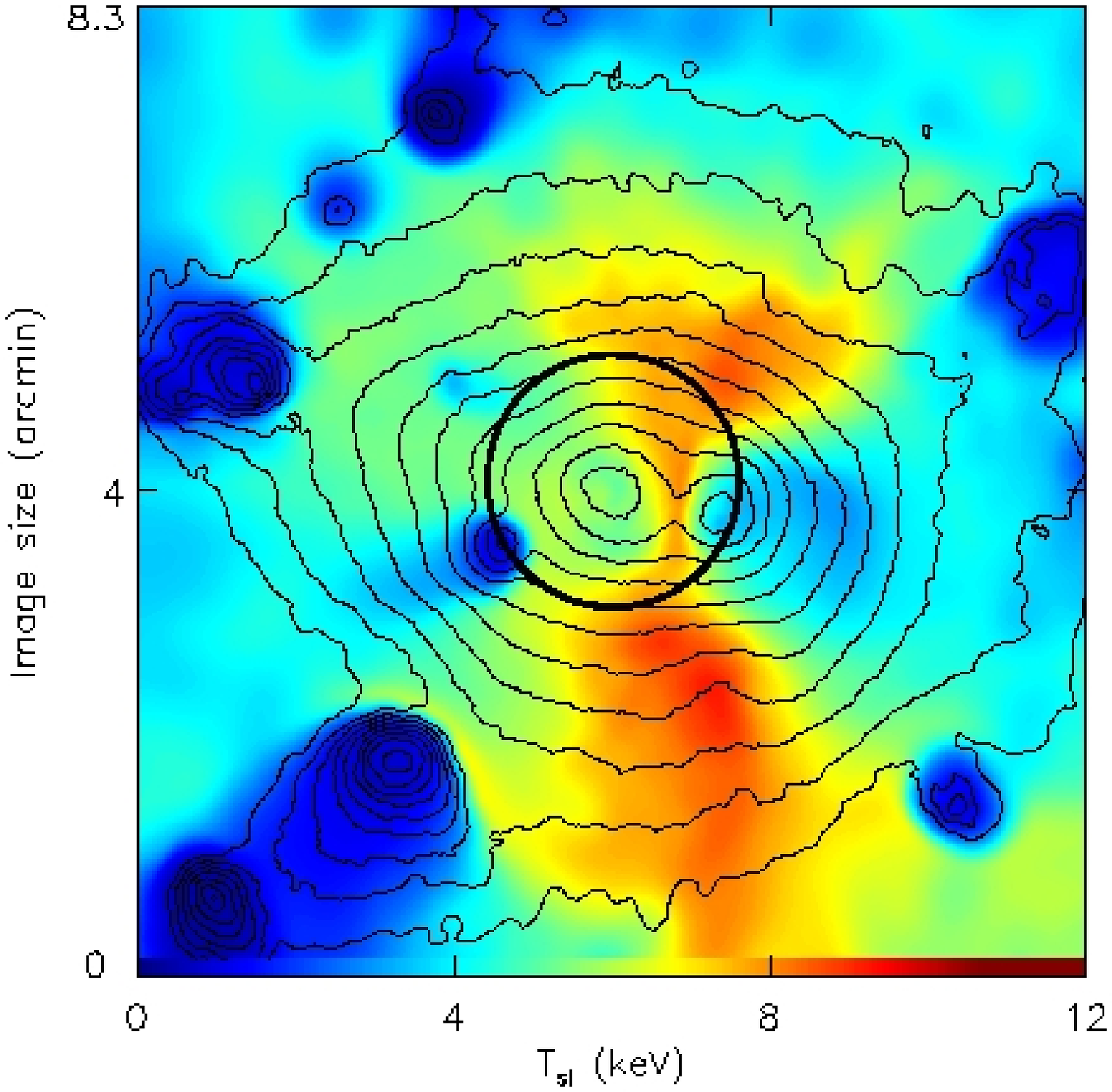,width=.25\textwidth}
\caption{Comparison of the map  of the
emission-weighted temperature \tew ~ (left-panel) and the
spectroscopic-like temperature \tsl ~ (right-panel) for the simulated
cluster of galaxy.  Both maps are obtained by using the gas particles
of the hydro-N-body simulation and are binned in $1''$ pixels.
The contours  correspond to the cluster flux.}
\label{fig:com}
\end{figure*}
\begin{figure*}
{\centering \leavevmode
\hfil \psfig{file=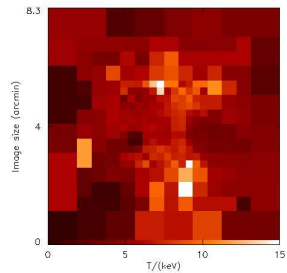,width=.25\textwidth}
}
\caption{Spectroscopic temperature map of the simulated cluster of galaxy as derived from the spectroscopic
analysis of the \chandra ~ ``observation'' with the package X-MAS}
\label{fig:comparison}
\end{figure*}

\begin{figure*}
\psfig{file=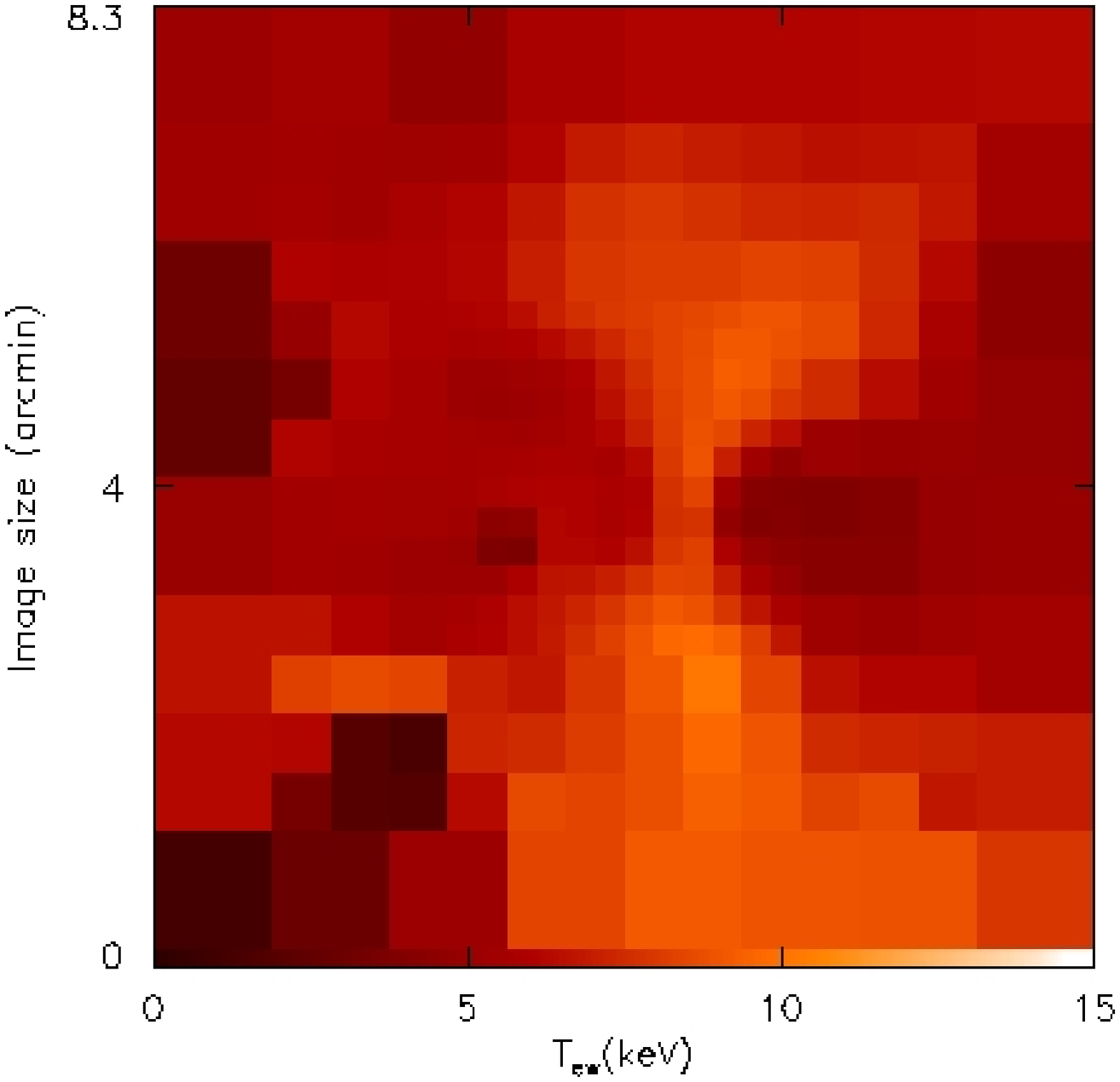,width=.25\textwidth} \hfil
\psfig{file=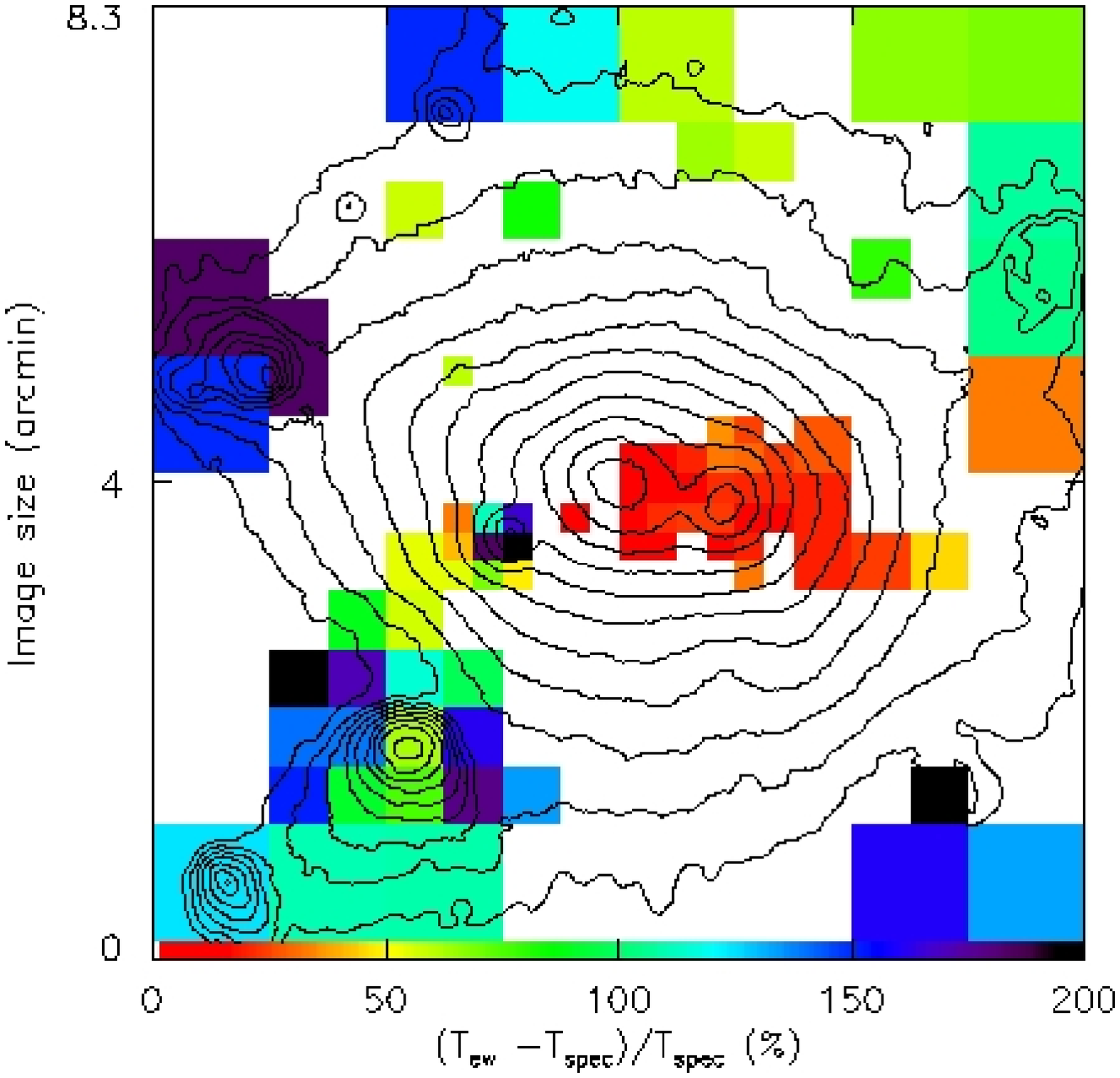,width=.25\textwidth} 
\caption{Left panel: emission-weighted temperature map of the simulated
cluster of galaxies shown in the left panel of Fig.~\ref{fig:com} but
re-binned to match the spatial resolution of the spectroscopic
temperature map shown in Fig.~\ref{fig:comparison}.  Right panel:
percentile difference between the spectroscopic and emission-weighted
temperature maps.  In this map we show only the regions where the
significance level of the temperature discrepancy is at least $3
\sigma$, i.e.  $|(T_{\rm ew}-T_{\rm spec}) /\sigma_{\rm spec}|>3$,
where $\sigma_{\rm spec}$ is the 68 per cent confidence level error
associated to $T_{\rm spec}$.}
\label{fig:comparison2}
\end{figure*}

 
A theoretical interpretation of these observations clearly requires state-of-the-art hydro-N-body
simulations, which can be used to extract realistic temperature maps
and/or profiles.  The comparison between real and simulated data,
however, is complicated by different problems, produced both by
projection effects and by instrumental artifacts.
 A further
complication can arise from a possible mismatch between the
spectroscopic temperature $T_{\rm spec}$ estimated from X-ray
observations and the temperatures usually defined in numerical
results.  In fact, while $T_{\rm spec}$ is a mean projected
temperature obtained by fitting a single or multi-temperature thermal
model to the observed photon spectrum, theoretical models fully
exploit the three-dimensional thermal information carried by gas
particles and so usually define physical temperatures.
In the years a number of temperature functions have been introduced in
order to compare hydro-N-body simulations with X-ray observations.
The most common temperature function used  is the so called bolometric
emission-weighted temperature function defined as
\be
T_{\rm ew}\equiv \frac{\int \Lambda(T) n2  T dV} {\int \Lambda(T) n2 dV}\ ,
\ee
where  $\Lambda(T)=\int_0^\infty \epsilon_E dE \propto \sqrt
{T}$ (see, e.g.,
[\citealt{1999ApJ...525..554F}]).
To make observations more directly comparable with simulations
we build the software package X-ray Map Simulator (X-MAS)
devoted to simulate X-ray observations of galaxy clusters obtained from hydro-N-body simulations
[\citealt{2003astro.ph.10844G}]. Using  X-MAS we clearly show that if the
cluster  has a complex thermal structure the emission-weighted
temperature function does not reproduce the spectroscopic estimate and
in particular it overestimates this value[\citealt{2003astro.ph.10844G}].
This problem is  illustrated in Fig.~\ref{fig:com},  Fig.~\ref{fig:comparison}, Fig.~\ref{fig:comparison2}.
In fact, in the  left panel of Fig~\ref{fig:com} we show the emission-weighted
temperature map
of a simulated cluster that we use as input for X-MAS.
We point to the reader the presence in this map of two
shock fronts:  one in the lower-left corner and the other at the center of the
Eastern side of the cluster.
In Fig.~\ref{fig:comparison} we show the \tspec ~ map
derived from the data analysis of a 300~ks \chandra ~ ACIS-S3 X-MAS
``observation'' of the cluster shown in Fig~\ref{fig:com} (see  [\citet{2003astro.ph.10844G}] for details).

%
%

In order to make a direct comparison of the ``observed'' temperature
map to the emission-weighted one, we decreased the resolution
of the latter to match the resolution of the former. Thus, in the left
panel
of Fig.~\ref{fig:comparison2} we report the same map
of $T_{\rm ew}$ shown on the left panel of Fig.~\ref{fig:com}, but
re-binned as the map of Fig.~\ref{fig:comparison}.
To highlight the temperature differences between the \tspec ~ and
the \tew ~ maps, in the right panel of
Fig.~\ref{fig:comparison2} we show 
$(T_{\rm ew}-T_{\rm spec})/T_{\rm spec}$. We
only show the pixels where this difference is significant to at least
$3 \sigma$ confidence level, i.e. $|(T_{\rm ew}-T_{\rm
spec})/\sigma_{\rm spec}|>3 $. This plot clearly shows that there are many
regions where the difference between $T_{\rm ew}$ and $T_{\rm spec}$
is significant to better than $ 3\sigma$ confidence level.
Furthermore, for these pixels the discrepancy ranges from 50 per cent
to 200 per cent and even more. Of particular relevance are two cluster
regions showing a shock front in the \tew ~ map, in the
left panel of Fig.~\ref{fig:com}: we notice discrepancies of 100-200
per cent, indicating that shock fronts predicted in the
\tew ~ map are no longer detected in the observed the \tspec ~ map.

%
%


%
%

\section{A new formula to estimate  spectroscopic temperatures}\label{par:slike}

If the cluster thermal structure is
complex, the $T_{\rm ew}$ may substantially differ from $T_{\rm spec}$.
This is mainly related to the fact that  the projected
temperature function is not a good physical quantity.
From a pure analytic point of view, in fact, the sum
of the spectra of two or
more thermal model  is no longer a thermal model [\cite{2004astro.ph..4425M}].
In the real world, however, the finite energy response and limited energy resolution of the X-ray
instruments 
conspire  to distort  the observed spectra
making multi-temperature thermal source spectra
fitted by single-temperature thermal models which have little to do
with the real temperature, but nevertheless are statistically
indistinguishable from it (see [\cite{2004astro.ph..4425M}] for details).
In [\cite{2004astro.ph..4425M}] we show that the temperature that better
approximates $T_{\rm spec}$  is given by what we call the
``spectroscopic-like" temperature $T_{\rm sl}$.
The idea behind  the derivation of this temperature function is quite
simple: if we assume two thermal components
with densities $n_1$, $n_2$,
and temperatures $T_1$, $T_2$, respectively, requiring matching
spectra means that
\be
\begin{array}{l}
n_{1}^2\zeta(Z,T_1)\frac{1}{\sqrt T_1}\exp(-\frac{E}{kT_1})+
n_{2}^2\zeta(Z,T_2)\frac{1}{\sqrt T_2}\exp(-\frac{E}{kT_2})\\
\approx A\zeta(Z,T_{\rm sl})\frac{1}{\sqrt T_{\rm
   sl}}\exp(-\frac{E}{kT_{\rm sl}}),\\
\end{array}
\label{eq:condizion}
\ee where $A$ is an arbitrary normalization constant and $\zeta(Z,T)$ is a
parametrization function that accounts for the total Gaunt factor and
partly for the line emission.

\begin{figure*}
\psfig{file=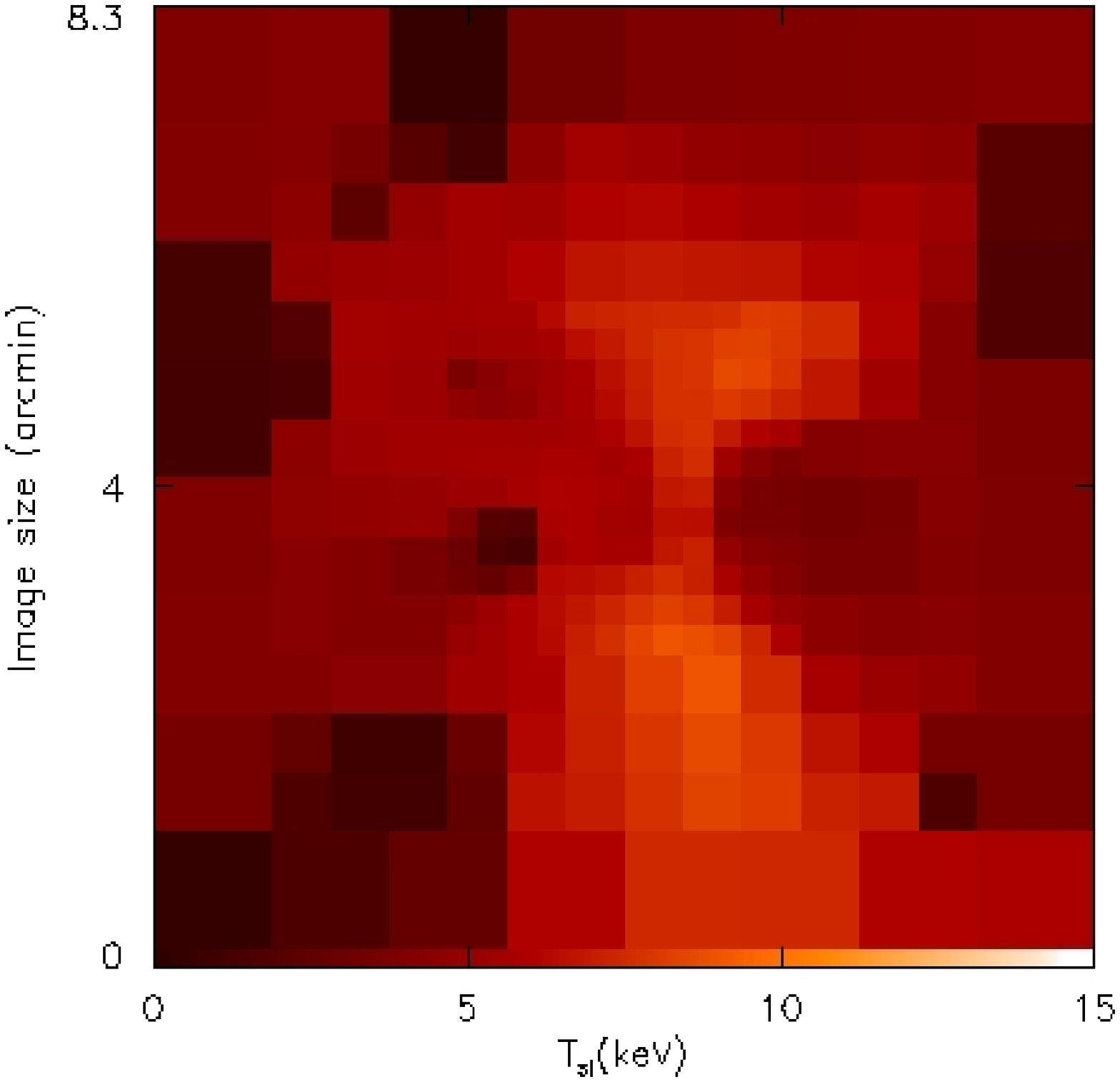,width=.25\textwidth} \hfil
\psfig{file=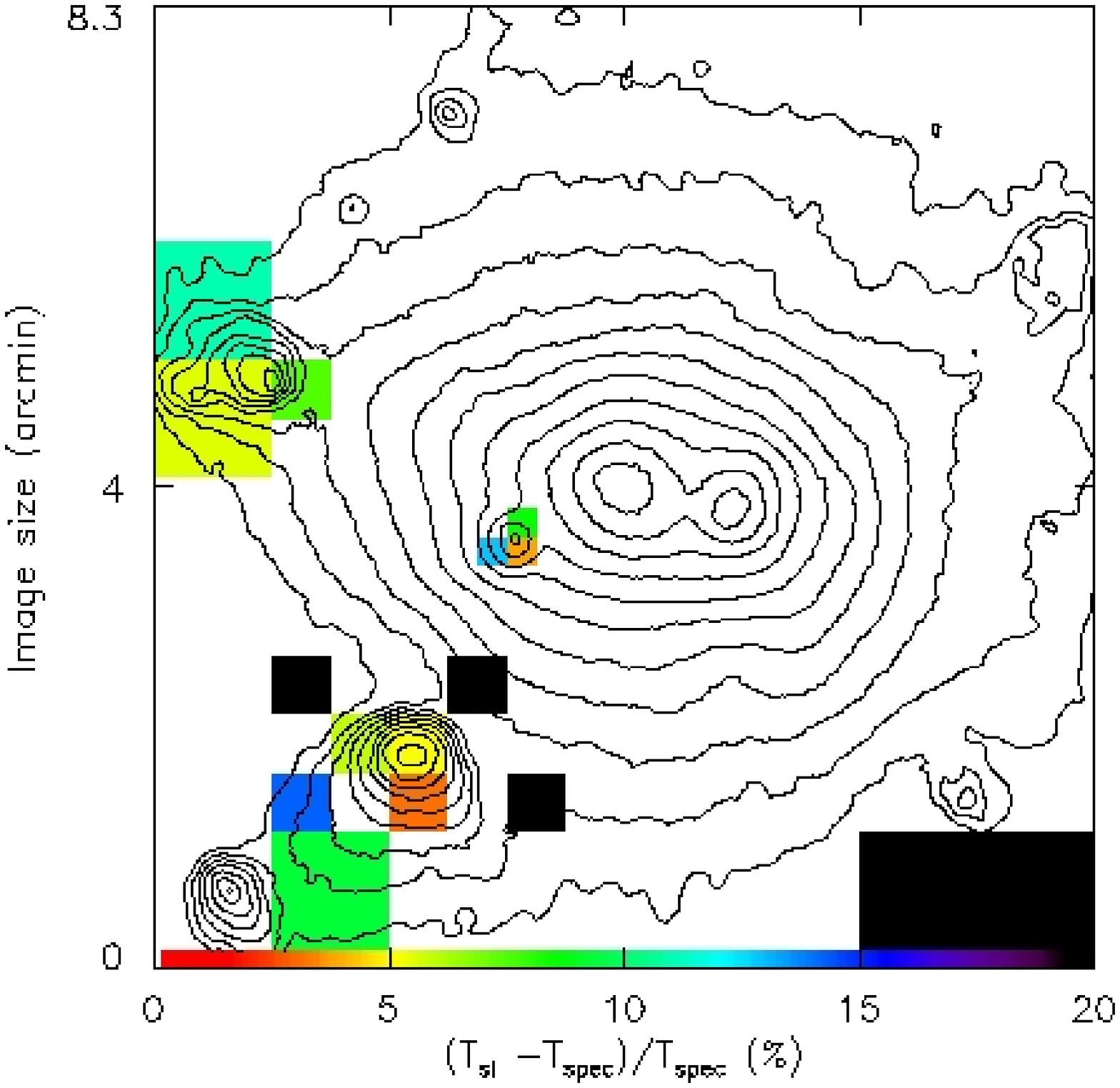,width=.25\textwidth}
\caption{As Fig. ~\ref{fig:comparison2}, but for the
spectroscopic-like temperature.
}
\label{fig:comparison3}
\end{figure*}

Both \chandra ~ and \xmm ~ are most sensitive to the soft region
of the X-ray spectrum, so we  expand both sides of
Eq.~\ref{eq:condizion} in Taylor series, to the first order in $E/kT$.
By  equating the zero-th and first-order terms in $E$, assuming
a power-law functional form for $\zeta(Z,T_2)$, and extending to a continuum
distribution we find that (see [\cite{2004astro.ph..4425M}] for details):
\be
\begin{array}{rl}
T_{\rm sl}&= \frac{\int W T dV }{\int W dV}, \\
W&=\frac{n2}{T^{3/4}}.\\
\end{array}
\label{eq:new_form_TW} \ee It is interesting to note that $T_{\rm sl}$ weights each thermal
component directly by the emission measure but, unlike \tew, inversely
by their temperature to the power of $3/4$.  This means that the observed
spectroscopic temperature is biased toward the coolest
regions.
To prove the quality of \tsl ~ in reproducing \tspec ~ we do  the same
test done for \tew ~ and shown in Fig.~\ref{fig:comparison2}. In the left panel of Fig.~\ref{fig:comparison3} we report
the cluster $T_{\rm sl}$  map re-binned as the map of Fig.~\ref{fig:comparison}.
In the right panel of Fig.~\ref{fig:comparison3} we
show  $(T_{\rm sl}-T_{\rm spec})/T_{\rm
spec}$.  The presence
in this map of fewer pixels clearly indicates that the match between
\tsl ~and \tspec ~ is much better than the one between \tspec ~
and \tew. Furthermore, most of these pixels shows very small
temperature discrepancies.
This demonstrates that the \tsl ~ temperature gives a much
better estimate of the observed \tspec than the
widely used \tew .

It is very important to say that, unlike the emission-weighted, the
map on the right panel of Fig.~\ref{fig:comparison3} does not show big
discrepancies between \tsl ~ and \tspec ~ in both shock
regions. This clearly indicate that \tsl ~ does a much better job than
\tew ~ in predicting the projected spectral properties of such peculiar
thermal features.

%
%

\section{Conclusions}\label{par:dis}

The projected \tspec ~  of
thermally complex clusters  is
lower than \tew .
This bias has important implications for the study of
thermal structures in clusters like shock fronts.  In fact in
real observations shock fronts appear much weaker than what is
predicted by emission-weighted temperature maps, and may even not be
detected.  This bias effect  is evident  on the right panel of
Fig.~\ref{fig:com} where we report the spectroscopic-like temperature map of
the simulated cluster. We notice that
the   $T_{\rm sl}$ map appears
cooler than the map of $T_{\rm ew}$. Furthermore,
 both shock fronts, which are
clearly evident in the \tew ~ map, are no
longer detected in the $T_{\rm sl}$ map.
This bias may explain why, although numerical simulations predict
that shock fronts are a quite common feature in clusters of galaxies,
to date there are very few observations of objects in which they are
clearly seen.
To conclude we stress that the emission-weighted temperature
function may give a misleading view of the actual gas temperature
structure as obtained from X-ray observations. Thus, here we propose to
theoreticians 
to finally discard its use. 




\end{document}